# Phase transitions study of the liquid crystal DIO with a ferroelectric nematic, a nematic and an intermediate phase and of mixtures with the ferroelectric nematic compound RM734 by adiabatic scanning calorimetry

J. Thoen[1*], G. Cordoyiannis[2], W Jiang[3], G. H. Mehl[3], C. Glorieux[1]

[1] Laboratory for Soft Matter and Biophysics, Department of Physics and Astronomy, KU Leuven, Celestijnenlaan 200D, 3001 Leuven, Belgium.

[2] Condensed Matter Physics Department, Jožef Stefan Institute, 1000 Ljubljana, Slovenia.

[3] Department of Chemistry, University of Hull, Hull HU6 7RX, UK.

*Corresponding author:* jan.thoen@kuleuven.be (Jan Thoen)

## Abstract

High-resolution calorimetry has played a significant role in providing detailed information on phase transitions in liquid crystals. In particular adiabatic scanning calorimetry (ASC), capable of providing simultaneous information on the temperature dependence of the specific enthalpy $h(T)$ and on the specific heat capacity $c_p(T)$, has proven to be an important tool to determine the order of transitions and render high-resolution information on pretransitional thermal behavior. Here we report on ASC results on the compound 2,3',4',5'-tetrafluoro[1,1'-biphenyl]-4-yl-2,6-difluoro-4-(5-propyl-1,3-dioxan-2-yl) benzoate (DIO) and on mixtures with 4-[(4-nitrophenoxy)carbonyl]phenyl 2,4-dimethoxybenzoate (RM734). Both compounds exhibit a low-temperature ferroelectric nematic phase ($N_F$) and a high-temperature paraelectric nematic phase (N). However, in DIO these two phases are separated by an intermediate phase ($N_x$). From the detailed data of $h(T)$ and $c_p(T)$, we found that the intermediate phase was present in all the mixtures over the complete composition range, albeit with strongly decreasing temperature width for that phase with decreasing mole fraction of DIO ($x_{DIO}$). The $x_{DIO}$ dependence of the transition temperatures for both the transitions could be well described by a quadratic function. Both these transitions were weakly first-order. The true latent heat of the $N_x$-N transition of DIO was as low as $L = 0.0075 \pm 0.0005$ J/g and $L = 0.23 \pm 0.03$ J/g for the $N_F$-$N_x$ transition, which is about twice the previously reported value of 0.115 J/g for the $N_F$-N transition in RM734. In the mixtures both transition latent heats decrease gradually with decreasing $x_{DIO}$. At all the $N_x$-N transitions pretransition fluctuation effects are absent and these transitions are purely but very weakly first order. As in RM734 the transition from the $N_F$ to the higher temperature phase exhibits substantial pretransitional behavior, in particular in the high-temperature phase. Power law analysis of $c_p(T)$ resulted in an effective critical exponent $\alpha = 0.88 \pm 0.1$ for DIO and this value decreased in the mixtures with decreasing $x_{DIO}$ towards $\alpha = 0.50 \pm 0.05$ reported for RM734. Ideal mixture analysis of the phase diagram was consistent with ideal mixture behavior provided the total transition enthalpy change was used in the analysis.

## 1. INTRODUCTION

Thermotropic liquid crystals are materials characterized by different degrees of orientational and partial positional order between crystalline solids and the isotropic liquid state [1]. The simplest liquid crystal phase is the nematic phase, as the least ordered phase, where rod-like or disk-like molecules are statistically oriented along a common direction, called the director, whereas the centers of mass of the molecules are randomly distributed. Depending on the shape of molecules or molecular aggregates, on the absence or presence of dipoles, on temperature and on mixture composition, different types of nematic phases are encountered. For example, phase transitions between uniaxial and biaxial nematic phases [2,3] as well as between two uniaxial nematic phases with different types of short-range smectic order [4] have been observed. Although Debye [5] and Born [6] already discussed more than a century ago the possibility of orientational order of polar rod-shaped molecules of a nematic phase with interacting molecular dipoles in such a way that the nematic phase is ferroelectric with a nonzero polarization density, such a phase has never been observed until recently. In 2017 two independent reports by Mandle *et al.* [7] and Nishikawa *et al.* [8] claimed the discovery of a new type of nematic phase for two quite different types of molecules. The two compounds that were subsequently extensively studied [8-18] are: RM734 (4-[(4-nitrophenoxy)carbonyl]phenyl 2,4-dimethoxybenzoate) [7] and DIO (2,3',4',5'-tetrafluoro[1,1'-biphenyl]-4-yl-2,6-difluoro-4-(5-propyl-1,3-dioxan-2-yl) benzoate) [8]. The chemical structures of DIO and RM734 are given in Fig. 1 The strongly polar compound RM734, exhibits two nematic phases separated by a phase transition, which was claimed to be weakly first order [13]. In 2020 Chen et al. [14] presented experimental evidence for ferroelectricity in the low temperature nematic phase ($N_F$) of RM734, the high temperature one being a normal uniaxial nematic phase (N). The compound DIO is special in the sense that in addition to the 'ferroelectric-like' [8] $N_F$ phase, it exhibits at low temperatures an additional phase between the $N_F$ phase and the normal nematic phase N. During the past two years the quest for new liquid crystals with ferroelectric nematic phases and their investigations gained substantial momentum. Efforts have largely concentrated on synthesizing variants of the RM734 and DIO families, to the extent that presently more than 50 molecules are known to induce ferroelectricity [19-22].

The true nature and proper naming of the intermediate phase is presently a matter of debate. In [8] it was initially called $M_2$ and subsequently $N_x$ in [17]. In a paper by Chen *et al*. [23] it was claimed, on the basis of structural characterization, that the intermediate phase of DIO ($M_2$ in [8] and $N_x$ in [17]), between the high-temperature (polar-disordered) dielectric nematic phase and the low temperature (polar-ordered) ferroelectric nematic phase $N_F$, is an antiferroelectric smectic phase with the nematic director and the polarization oriented parallel to smectic layer planes, and the polarization alternating in sign from layer to layer. This phase was termed smectic $Z_A$ and indicated as $SmZ_A$. However, in a very recent paper Sebastian *et al.* [24] argued that the intermediate phase of DIO should be considered as a splay nematic phase and be named $N_S$. Because calorimetry cannot render information on the structure of phases and in view of the above variation in naming of the intermediate phase, we will, further in this paper, use the terminology $N_x$ to indicate this intermediate phase between the N and $N_F$ phases. In another recent paper, Chen *et al.* [25] reported on experimental investigations of the phase diagram and electro-optics of binary mixtures of RM734 and DIO, and found complete miscibility in the dielectric nematic as well as in the ferroelectric nematic phases.

---

In recent high-resolution adiabatic scanning calorimetric measurements of the $N_F$-N transition of RM734, reported by Thoen *et al.* [26], the order of the transition as well as the pretransitional fluctuation-induced critical behavior was investigated in detail. The $N_F$-N transition was found to be very weakly first order with a latent heat $L = 0.115 \pm 0.005$ J/g (which is more than an order of magnitude smaller than the values derived from different scanning calorimetry (DSC) [13.14]). In both the N and $N_F$ phases the power-law analysis of the specific heat capacity, $c_p$, resulted in an effective critical exponent $\alpha = 0.50 \pm 0.05$ and an amplitude ratio $A_{N_F}/A_N = 0.42 \pm 0.03$. The very small latent heat and the value of $\alpha$ indicate the $N_F$-N transition in RM734 to be close to a tricritical point. This conclusion was additionally supported by an order parameter critical exponent value of $\beta \approx 0.25$ obtained from electric polarization measurements in the $N_F$ phase (Ref. [32] in [26]).

DSC measurements have indicated the $N_F$-$N_x$ and the $N_x$-N transitions of DIO to be first order. However, it should be realized that it is inherently difficult for DSC to distinguish between true latent heats and pretransitional fluctuations-induced enthalpy increases [27-29]. In view of that, we present here results for DIO and for its mixtures with RM734 of simultaneous measurements of the temperature dependence of the specific heat capacity $c_p(T)$ and of the specific enthalpy $h(T)$ by high-resolution adiabatic scanning calorimetry (ASC). The high-resolution data of $c_p(T)$ and $h(T)$ near both phase transitions, $N_F$-$N_x$ and $N_x$-N, will be used to determine the order of the transitions and characterize the pretransitional critical behavior.

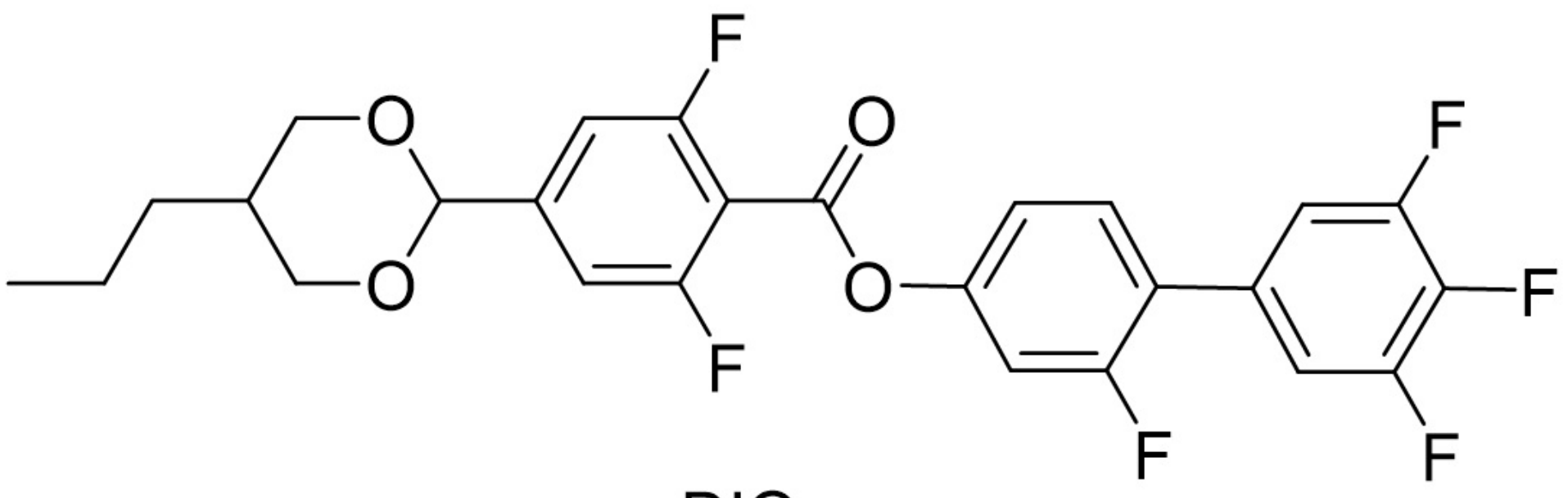

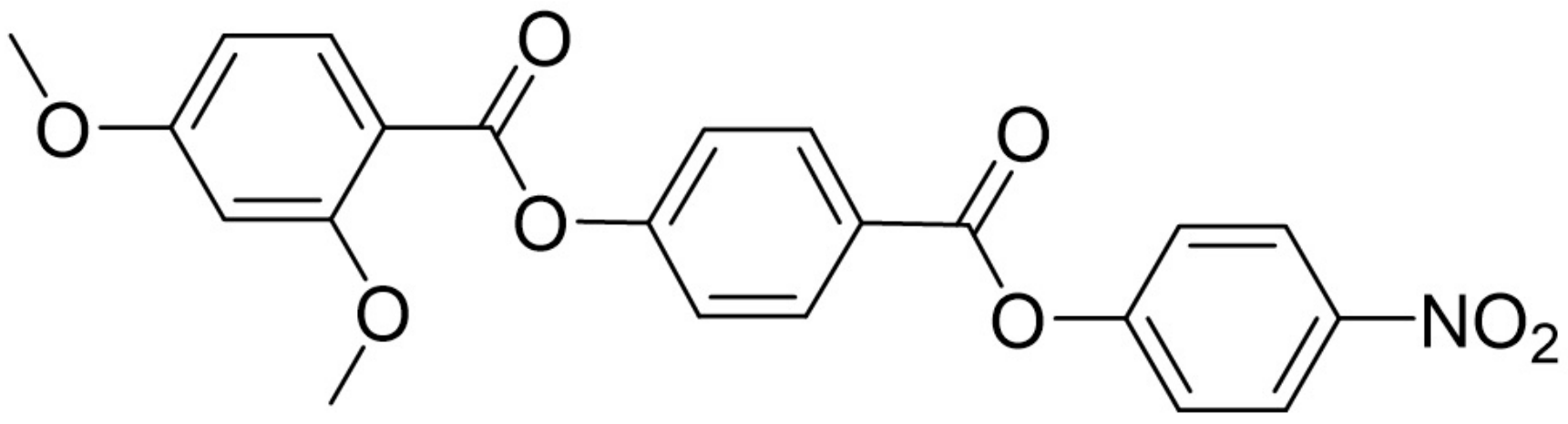

Fig. 1 Chemical structure of DIO (top) and RM734 (bottom).

## 2. EXPERIMENTAL

### *2.1 Adiabatic scanning calorimetry (ASC)*

#### *2.1.1 Operational principle.*

ASC is used to obtain simultaneously and continuously the temperature evolution of the heat capacity $C_p$ and the enthalpy $H$ of a sample under investigation [29-33]. The basic concept of ASC is in applying a constant heating or cooling power to a sample holder containing the sample. The sample holder is placed inside a surrounding adiabatic shield; in practice this is implemented by a number of shields and by continuously vacuum pumping the calorimeter. In a heating run the heat exchange between inner shield and sample holder is cancelled by keeping the temperature difference zero at all times. In a cooling run the heat exchange is controlled and monitored. During a run the sample temperature $T(t)$ is recorded as a function of time $t$. Together with the applied power $P$ this directly results in the enthalpy curve

$$H(T) - H(T_0) = \int_{t_0}^{t(T)} P dt = P[t(T) - t_0(T_0)], \quad (1)$$

where $H(T_0)$ is the enthalpy of the system at temperature $T_0$ at the starting time $t_0$ of the run. The heat capacity $C_p(T)$ is easily calculated via the ratio of the know constant power $P$ and the changing temperature rate $\dot{T} = dT/dt$,

$$C_p = \frac{P}{\dot{T}} \, . \quad (2)$$

The specific heat capacity $c_p(T)$ and the specific enthalpy $h(T)$ are obtained by using the sample mass and the calibrated background values of the empty calorimeter and of the used sample cells. It should be noted that keeping $P$ constant in Eq. (2) is completely opposite to the operation of a DSC where one imposes a constant scanning rate $\dot{T}$ on a sample and on a reference and measures the difference in heat flux, $\Delta P(t)$, between the sample and the reference.

*2.1.2 Peltier-element-based adiabatic scanning calorimeter (pASC)*

Maintaining adiabatic conditions over a long time and over wide temperature ranges as implemented in 'classical' versions of ASC is challenging. Detailed descriptions of this type of implementations and experimental results can be found in [29-33] and references therein. However, in the last decade this approach has been superseded by the development of the novel Peltier-element-based adiabatic scanning calorimeter (pASC), providing greater user-friendliness and allowing much wider temperature ranges as well as much smaller samples[34-38].

In a pASC the uncertainty on the final results, $c_p(T)$ and $h(T)$, depends on the uncertainties on the measured quantities $T(t)$, $P(t)$, the sample mass $m$ and the heat capacity of the addenda $C_{add}$. It is assumed that the error on the time, $t$, is negligible. The addenda of the calorimeter are the sample cell and the sample holder, consisting of a thin copper supporting platform with the heater, the thermistor and the top plate of the Peltier element. In order to assure good thermal contact between the bottom of the sample cell and the platform, a small amount (typically of the order of 1 mg) of thermal paste is applied. In extended separate runs of the calorimeter the heat capacity of the sample holder, as well as the specific heat capacity of the sample cell material and of the thermal paste have been evaluated. From an extended error analysis presented in reference [39] and its supplementary material it was concluded that a relative standard uncertainty of 2 % could be assigned to $c_p$ and $h$, provided the uncertainty on the sample mass is kept below this value.

### *2.1 Measurements and sample preparation*

Measurements have been carried out with the same pASC as the one used for the measurements on neat RM734 [26]. Samples (used as received) were transferred into stainless steel sample holders (Mettler Toledo 120 µl medium pressure crucibles) and hermetically sealed. These crucibles were vacuum-tight-closed with a miniature elastic O-ring between the cell body and a lid. Mixtures were prepared in the sample holders. The different masses were measured using a Mettler Toledo balance with a precision of 0.1 mg. The amounts of samples (typically 50 mg) used were chosen such that that the total uncertainty was within 1 %. In order to ensure proper mixing, the closed filled crucibles were placed on top of a hot plate at a temperature above the nematic-isotropic transition temperature. Subsequently, the crucible with the sample was positioned upside down on the hot plate for some time, and reversed to its normal position. This procedure was typically repeated 10 times, and took between 15 to 30 min.

## 3. RESULTS

### *3.1 Neat DIO*

#### *3.1.1 Melting transition.*

After a sample cell with a pure DIO sample was mounted in the calorimeter, the calorimeter was heated to 115 °C. This resulted in melting of the sample from the solid phase into the nematic phase via a sharp first-order transition at 95.8 °C. In Fig. 2 the temperature dependence of the specific enthalpy and of the corresponding effective specific heat capacity are displayed. A detailed analysis of the data resulted in a latent heat of 54.9 ± 0.3 J/g. The cusp in the enthalpy curve at the high temperature side of the transition is much sharper than on the low temperature side where some curvature is observed. This is a general aspect seen at many melting transitions [35] and which is in particular noticed in the melting of alkane compounds [38].

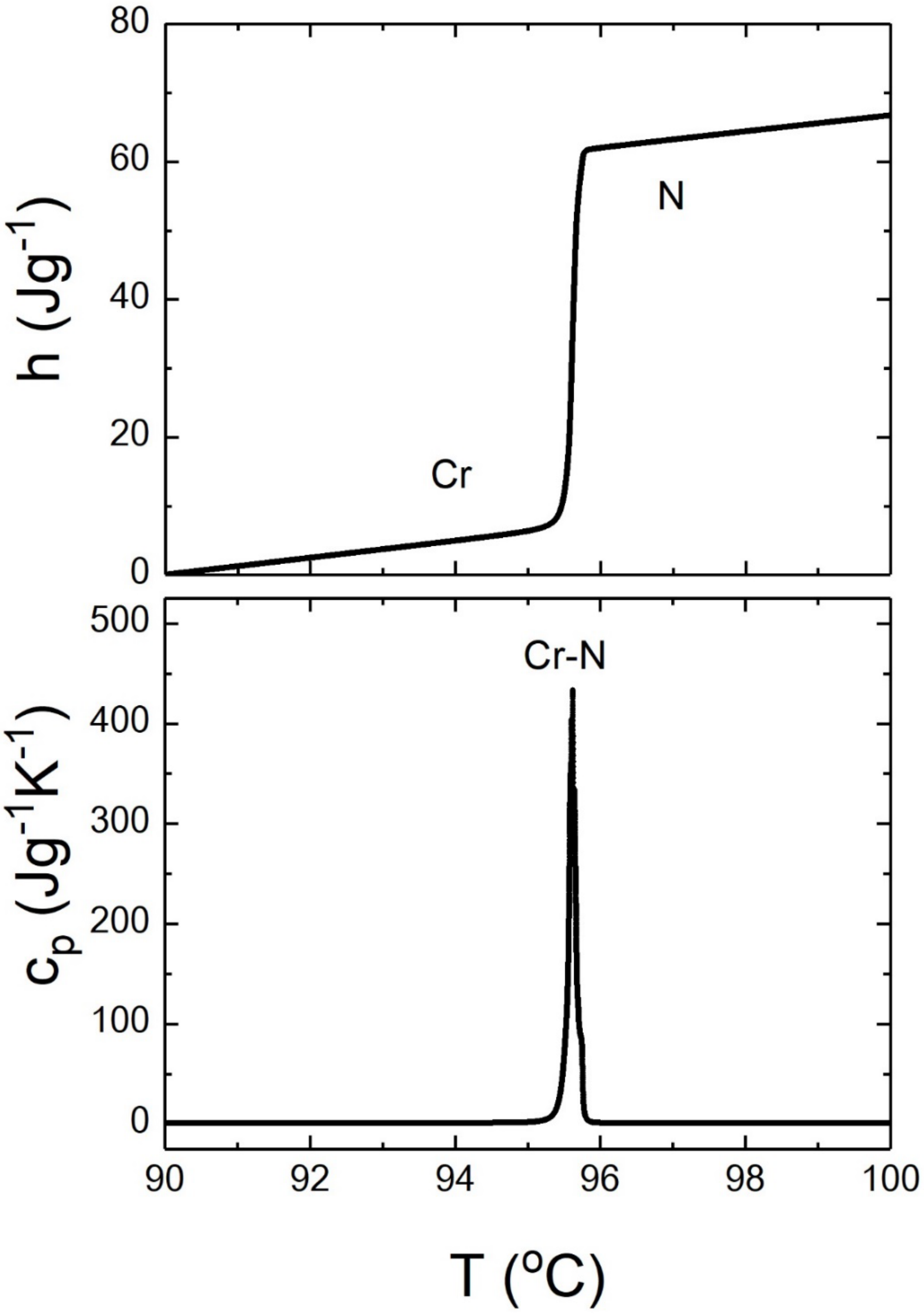

Fig. 2 Adiabatic scanning calorimetry results above and below the melting transition temperature of neat DIO, upon heating. Temperature dependence of the effective specific heat capacity $c_p(T)$ and of the specific enthalpy $h(T)$ of neat DIO across the first-order transition from the solid crystalline phase to the nematic phase.

A slow cooling run (at an average rate of – 0.04 K/min) was subsequently started at 115 °C. This resulted in clear observations of the transitions from the N to the $N_x$ and from the $N_x$ the $N_F$ phases. However, the sample solidified at a temperature of 59 °C. Remelting (by heating to a temperature above 100 °C) and subsequently cooling always resulted (at our slow scanning rates) to solidification between 55 °C and 60 °C. Cooling to a temperature of 60 °C allowed us to carry out slow heating runs from well into the $N_F$ phase to well into the paraelectric N phase. Several heating and cooling runs (at different values of the power $P$ in Eqs. (1) and (2) resulting in different average rates) were executed on different samples while the samples were in the subcooled monotropic phase. Good consistency regarding the $h(T)$ and $c_p(T)$ was observed between the results for different runs and samples. However, some mild and progressive downward shift of the transition temperatures was observed, in particular when the samples were kept for prolonged times (of the order of several days) at high $T$ in the N phase. The observed shifts were typically of a few tenths of a degree over the time scale of a week and not relevant for a specific heating or cooling run.

*3.1.2 Ferroelectric nematic $N_F$ to the intermediate $N_x$ transition.*

Fig. 3 gives an overview of the temperature dependence of the specific heat capacity $c_p(T)$ over the temperature range between 60 °C and 90 °C, covering both the transition from the ferronematic $N_F$ to the intermediate $N_x$ phase and from the $N_x$ to the nematic phase. In this overview figure the peak of the $N_x$-N transition is extremely small in comparison with the $N_F$-$N_x$ transition (but better visible in the inset). Fig. 4 gives an overview of the transition related specific enthalpy $h(T)$ variation over the $N_F$-$N_x$ transition. Detailed analysis of the $h(T)$ in combination with $c_p(T)$ revealed a small 0.12 ± 0.03 K two-phase region and a true latent heat of 0.23 ± 0.03 J/g, indicating the weakly first-order character of this transition. Above this transition (in the $N_x$ phase) substantial fluctuations induced enthalpy variation is observed. Such variation is substantially weaker in the $N_F$ phase. This indicates that the total, transition related enthalpy change is composed of two contributions: the true latent heat $L$ and the pretransitional part $\delta h$. The total enthalpy change $\Delta h_{tr}$ over the transition (typically obtained in DSC) is the sum of both, thus:

$$\Delta h_{tr} = L + \delta h \qquad (3)$$

In order to obtain the transition enthalpy change $\Delta h_{tr}$, it is necessary to separate it from the enthalpy change that would occur without the transition. Indeed, when one has a constant background specific heat capacity $c_b$, the enthalpy increases linearly with temperature. In general $c_b$ can depend on temperature and usually a linear dependence is chosen (much like a baseline choice in DSC) between $c_p(T_1)$ and $c_p(T_2)$ well below and well above the transition, leading to

$$c_b\,(T) = c_p(T_1) + \frac{c_p(T_2)-c_p(T_1)}{T_2-T_1}(T-T_1). \qquad (4)$$

In this equation a linear temperature dependence is assumed, but not required. In some cases different choices are more appropriate, in particular when there are substantial differences in specific heat capacity background values between the low and high temperature phases [34]. Subtraction of $c_b(T)$ from the $c_p(T)$ peak and integration gives

$$\Delta h_{tr} = \int_{T_1}^{T_2} \left[c_p(T) - c_b(T)\right] dT. \tag{5}$$

Alternatively, one can use the direct specific enthalpy $h(T)$ data via

$$\Delta h_{tr} = h(T_2) - h(T_1) - \int_{T_1}^{T_2} c_b(T) dT \tag{6}$$

Using both approaches we obtained for $\Delta h_{tr}$ a value of 0.48 ± 0.03 J/g. In turn this resulted in $\delta h$ = 0.25 ± 0.03 J/g by subtracting the latent heat value. Applying the same procedures to the data of RM734 [26] resulted in $\Delta h_{tr}$ = 1.03 ± 0.1 J/g and in $\delta h$ = 0.91 ± 0.1 J/g. The choices made for the values $T_1$ and $T_2$ are to some extend arbitrary and they are very important only for small peaks. We typically use values of 5 K above and below the transition temperature. The quoted uncertainties reflect reasonable variations in these temperatures.

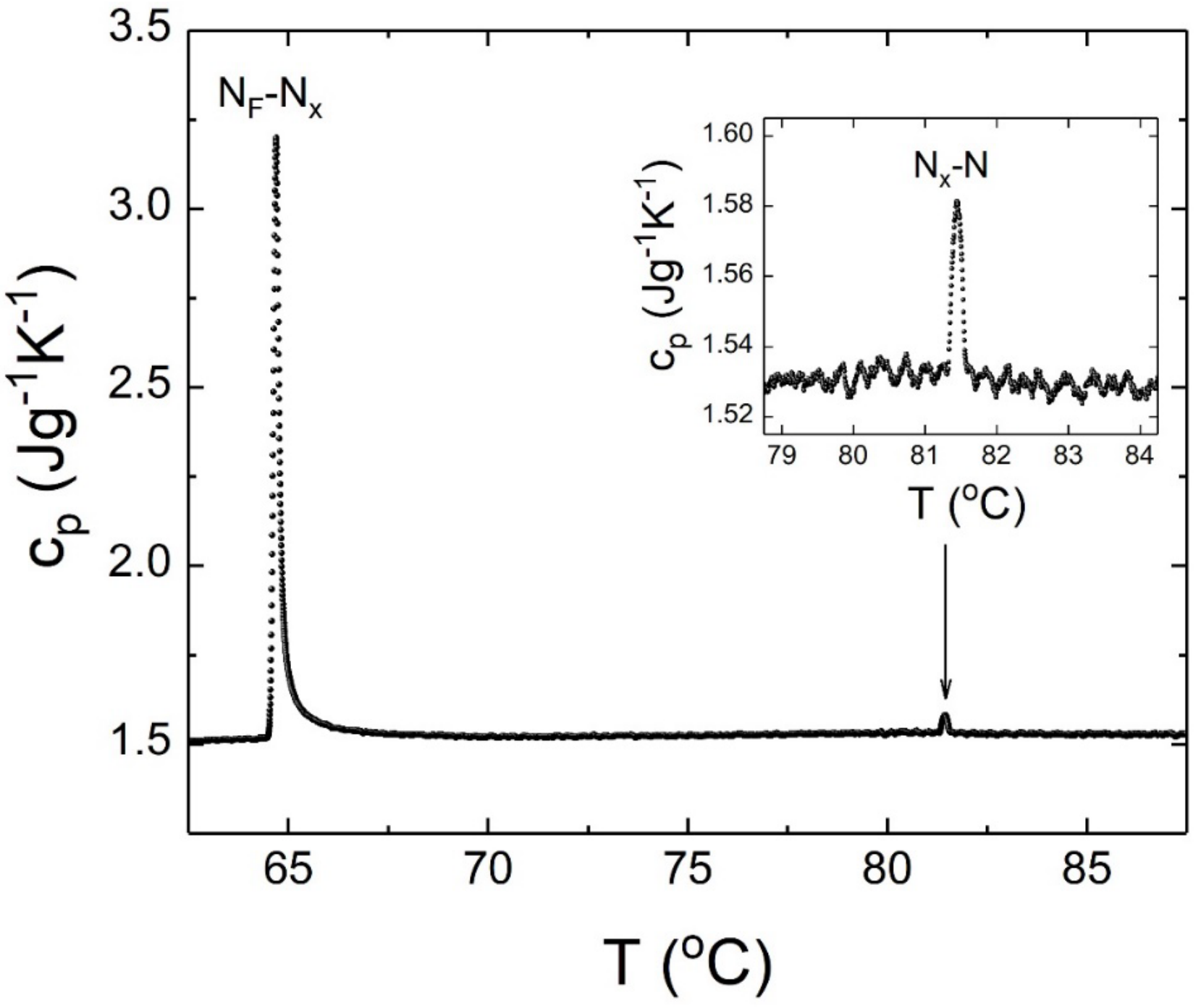

Fig. 3 Adiabatic scanning calorimetry results for neat DIO. Overview of the temperature dependence of the effective specific heat capacity $c_p(T)$ between 60 °C and 90 °C crossing the $N_F$-$N_x$ and the $N_x$-N transitions. The inset gives the detailed behavior around the $N_x$-N transition.

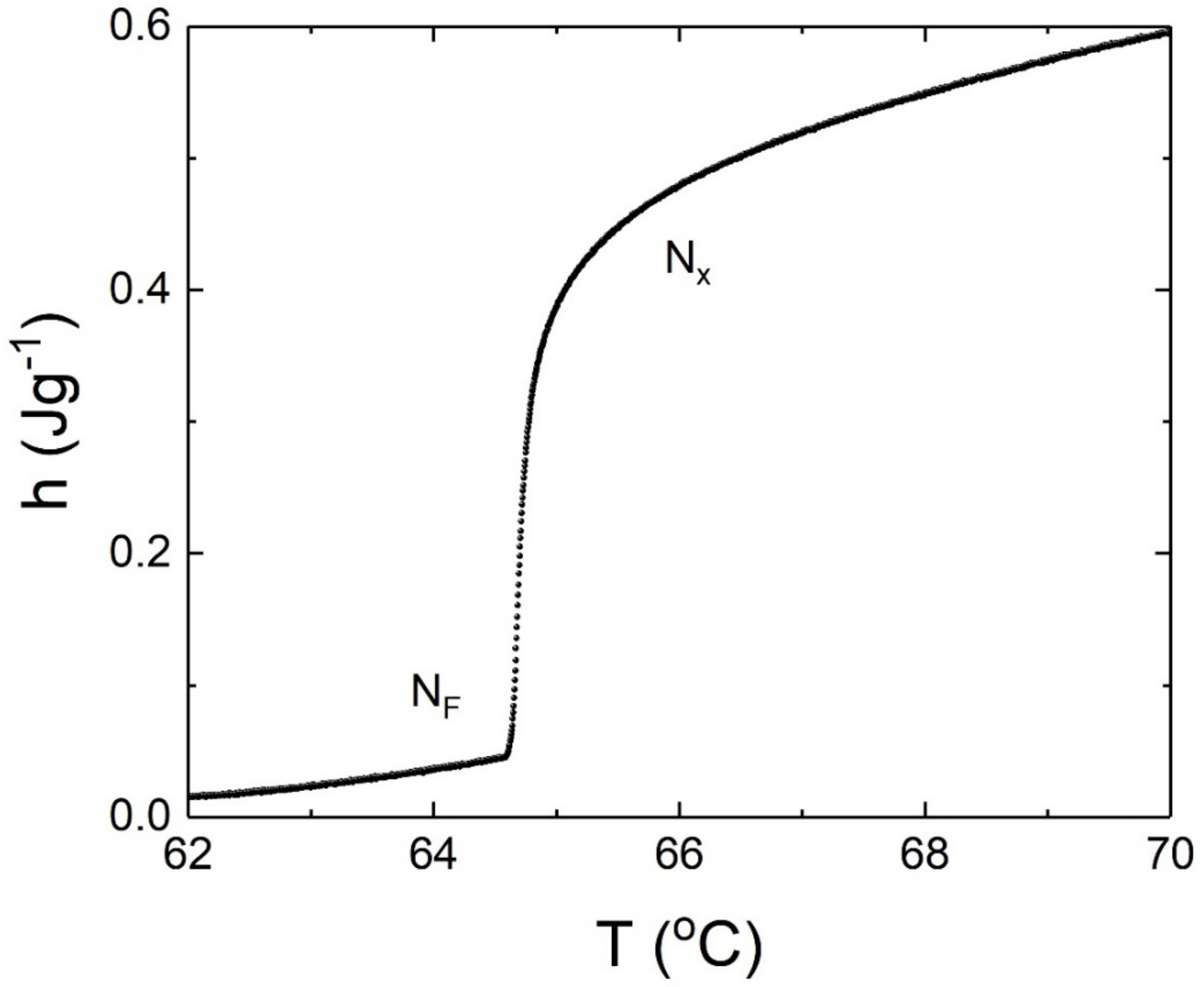

Fig. 4 Adiabatic scanning calorimetry results for neat DIO. Temperature dependence of the specific enthalpy $h(T)$ across the transition from the ferroelectric nematic $N_F$ phase to the $N_x$ phase after subtraction (for display reasons) of a linear temperature background $1.5(T - T_{ref})$ (J/g), with $T_{ref}$ an arbitrary reference temperature.

### *3.1.3 Intermediate $N_x$ to paranematic N transition*

As already observed in Fig 3, the enthalpy change associated with the transition between the $N_x$ and the N phase must be extremely small. Nevertheless ASC is quite capable to give a clear picture of $h(T)$ and of $c_p(T)$ around the transition. This can be seen in the detailed Fig. 5 in the temperature range between 80.5 °C and 83 °C. A clear first-order enthalpy discontinuity with a latent heat of 0.0075 ± 0.0005 J/g and a two-phase region of 0.11 ± 0.03 K are observed. The transition is purely first order without pretransitional fluctuation contributions.

### *3.1.4 Anomalous pretransitional specific heat capacity behavior at the $N_F$-$N_x$* transition.

Although the $N_F$-$N_x$ transition turns out to be weakly first order, from the data displayed in Figs. 3 and 4 it is obvious that substantial pretransitional fluctuations effects, in particular in the high-temperature phase, are present. In analogy with what can be done for purely second-order phase transitions, one can try to describe the pretransitional behavior of relevant physical quantities in terms of power laws with critical exponents, which depend on the universality class of the phase transition [40,41].

The limiting behavior of the specific heat capacity at a second-order phase transition can be described with a power law of the form

$$c_p = A|\tau|^{-\alpha} + B, \tag{7}$$

with $\tau = (T - T_c)/T_c$. $A$ is the critical amplitude, $\alpha$ is the critical exponent, $T_c$ is the critical temperature ($T$ and $T_c$ in kelvin) and $B$ is the background term. The different coefficients in Eq. (7) must be derived from (non-linear) least-squares fitting of experimental data. The fact that ASC scans result directly in enthalpy $h(T)$ data (see Eq. (1)), allows for a substantial simplification. One can introduce the following quantity:

$$C = \frac{h-h_c}{T-T_c}, \tag{8}$$

which corresponds to the slope of the chord connecting $h(T)$ at $T$, with $h_c$ at $T_c$. It can easily be shown that $C$ has a power law behavior of the form [29,31,32]:

$$C = \frac{A}{1-\alpha}|\tau|^{-\alpha} + B. \tag{9}$$

Both $c_p$ and $C$ have the same critical exponent, and either equation (7) or (9) can be used in fitting data to arrive at values for the critical exponent $\alpha$ and amplitude $A$. However, by considering the difference ($C - c_p$), above or below $T_c$, the (unimportant) background term $B$ drops out, resulting in:

$$C - c_p = \frac{\alpha A}{1-\alpha}|\tau|^{-\alpha}. \tag{10}$$

Taking the logarithm on both sides of Eq. (10) gives:

$$\log\left(C - c_p\right) = \log\left(\frac{\alpha A}{1-\alpha}\right) - \alpha \log|\tau|. \tag{11}$$

As a result, one obtains sufficiently close to the critical point, a straight line with a negative slope, immediately giving the critical exponent $\alpha$.

This procedure is strictly only applicable to second-order transitions, but for weakly first-order transitions it can be used for separate analysis of the data below and above the transition by allowing $T_c$ and $h_c$ in Eq. (8) to be adjustable parameters in fitting. The parameters $T_c$ and $h_c$ can be different for data below and above the transition. This is analogous to the upper stability limit of the nematic phase and the lower stability limit of the isotropic phase for the weakly first-order nematic-isotropic transition [31,41]. We have applied this approach to the present DIO phase transition data of $c_p$ and $h$, excluding the data in the two-phase region.

A double logarithmic plot (see Eq. (11)) is shown in Fig. 6 for $(C - c_p)$ data of two different samples (A and B) as a function of the reduced temperature difference $\tau$. It can be concluded that, within the experimental resolution, a negative slope of -0.88 ± 0.10 is consistent with the data. Thus, according to Eq. (11) this results in an effective critical exponent α = 0.88 ± 0.10. In Fig. 6, it can also be seen that $(C - c_p)$ above the transition $N_F$-$N_x$ is much larger than in the $N_F$ phase below the transition. From equation (7) it immediately follows that the critical amplitude in the ferroelectric nematic phase, $A_{N_F}$, must be substantially smaller than the critical amplitude in the $N_x$ phase. An estimate gives for the ratio $A_{N_F}/A_{N_x}$ = 0.39 ± 0.03.

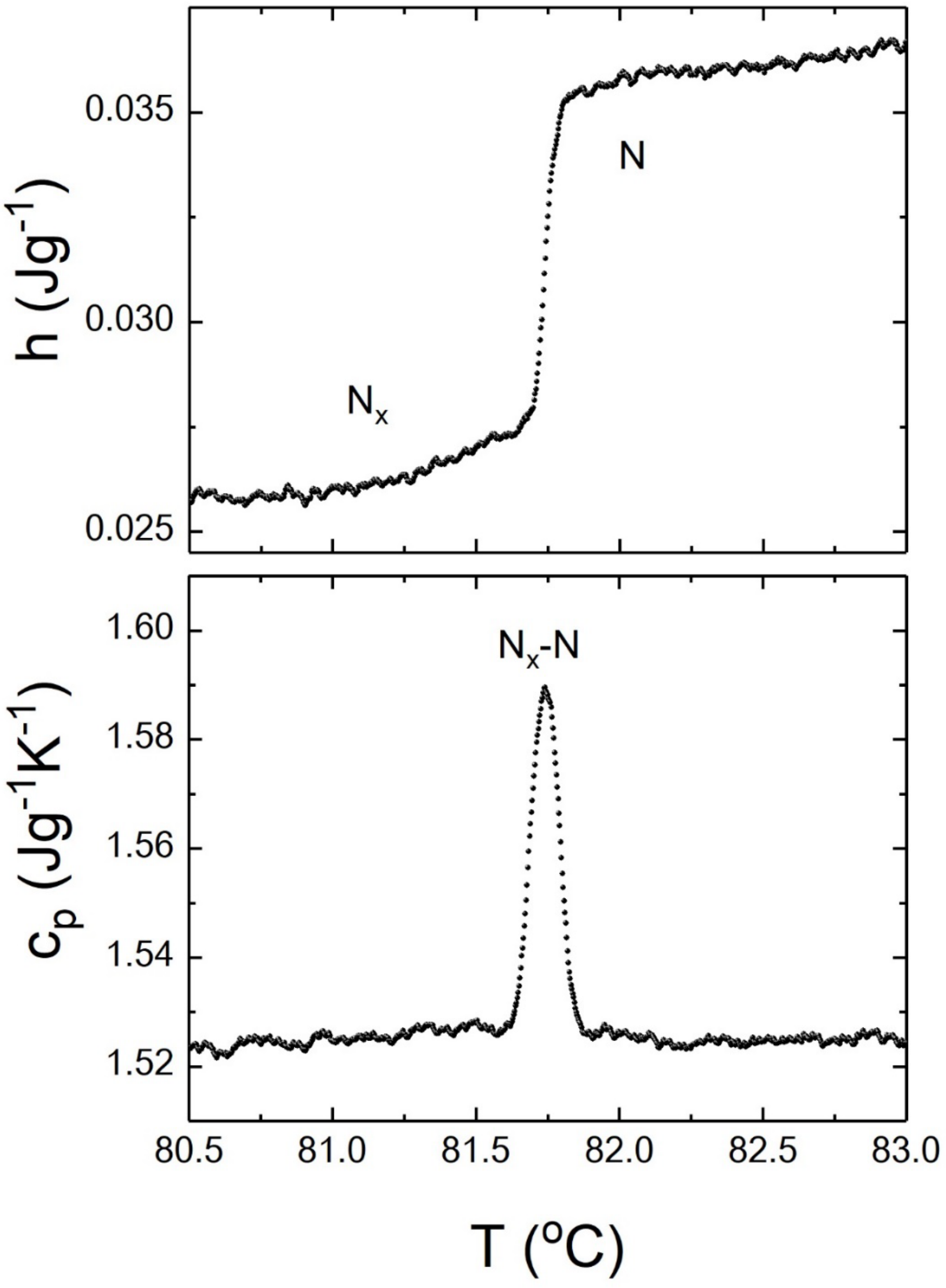

Fig. 5 Adiabatic scanning calorimetry results for neat DIO. The temperature dependence of the effective specific heat capacity $c_p(T)$ and of the specific enthalpy $h(T)$ is shown across the very small first-order transition from the $N_x$ phase to the nematic phase N. From the data of $h(T)$ a linear temperature dependent background $1.5(T - T_{ref})$ (J/g), with $T_{ref}$ an arbitrary reference temperature, has been subtracted.

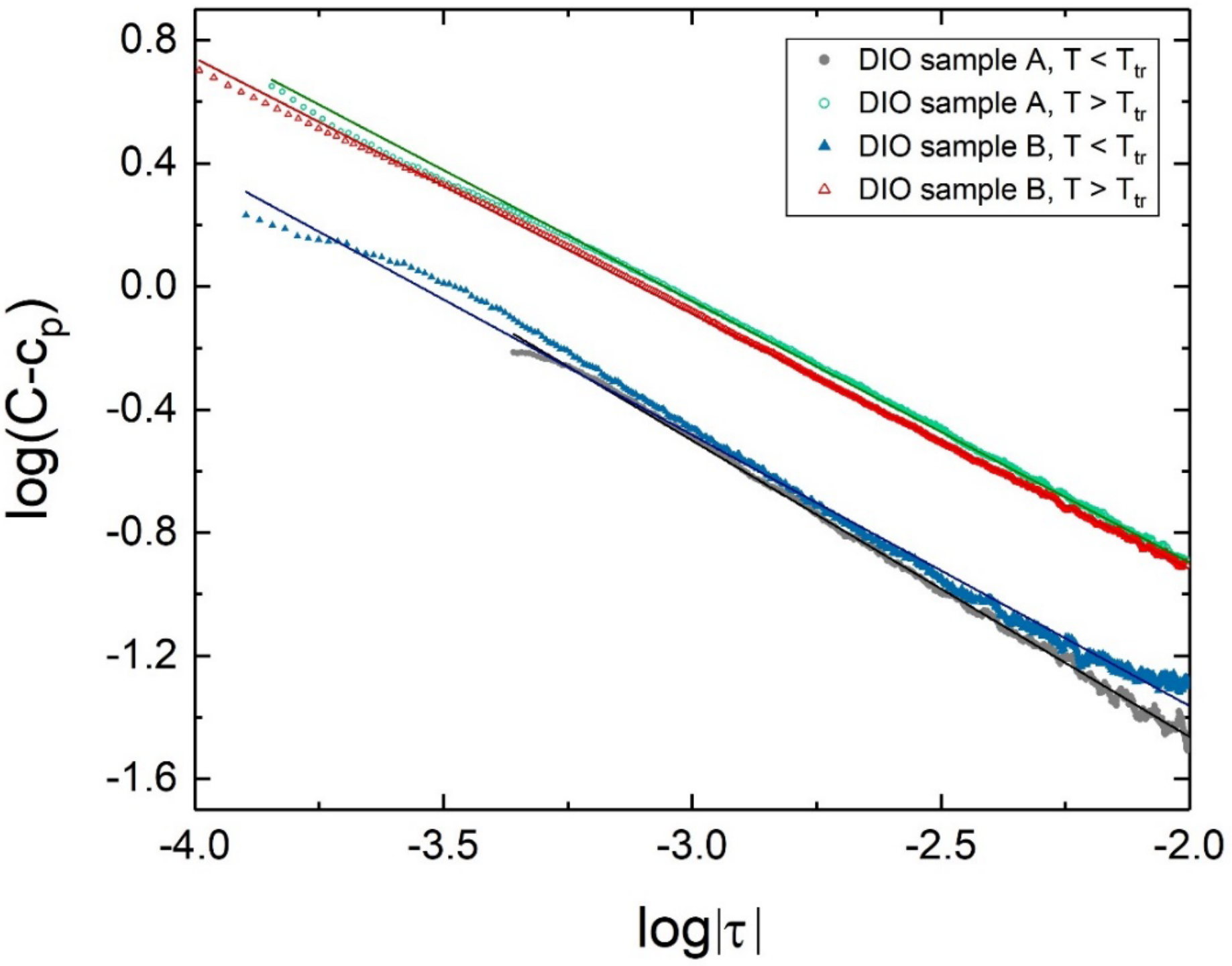

Fig. 6 (Color online). Adiabatic scanning calorimetry results for the $N_F$-$N_x$ transition for two different samples of neat DIO. Double logarithmic plot for the difference ($C - c_p$) expressed in J/(gK) as a function of the reduced temperature difference $\tau$. (see Eq 8).The upper open green circles (sample A) and open red triangles (sample B) are for the data of $T > T_{tr}$. The gray solid circles (sample A) and blue solid triangles (sample B) are for the data of $T < T_{tr}$. The average slopes of the different trend lines are consistent with an effective critical exponent $\alpha \approx 0.88 \pm 0.10$.

## *3.2 Mixtures of DIO and RM734*

### *3.2.1 Phase diagram*

In order to clearly establish the temperatures of the $N_F$-$N_x$ and the $N_x$-N transitions and the thermal behavior of the specific heat capacity $c_p(T)$ and of the specific enthalpy $h(T)$, eight different mixtures of DIO+RM734 were investigated by ASC. In Fig. 7 the temperature dependence of the specific heat capacity $c_p(T)$, from well in the ferroelectric nematic $N_F$ phase to well into the paraelectric nematic N phase over the $N_x$ phase, are compared for several DIO+RM734 mixtures. An overview of all the $N_F$-$N_x$ and $N_x$-N transition temperatures is depicted in Fig. 8. The plotted points correspond with the maximum of the $c_p(T)$ data. These results indicate that a narrowing of the $N_x$ phase occurs by reducing the amount of DIO in the mixture. For the lowest DIO weight fraction $w_D$= 0.118 investigated, the data still indicate a narrow T-range of 1.7 °C for a $N_x$ phase. This suggests the disappearance of this phase at $w_D$ = 0. This is in contradiction to the results in Ref. [25], where non-existence of the $N_x$ phase is reported for $w_D$ values below ~ 0.40. These latter data would mean a triple point around $w_D$ = 0.40, provided the two transition lines are first order ones. Moreover, contrary to Ref. [25] our results reveal a clear curvature in $w_D$ dependence of the transition temperatures for both transitions. This is clearly visible from the quadratic fits through the data points.

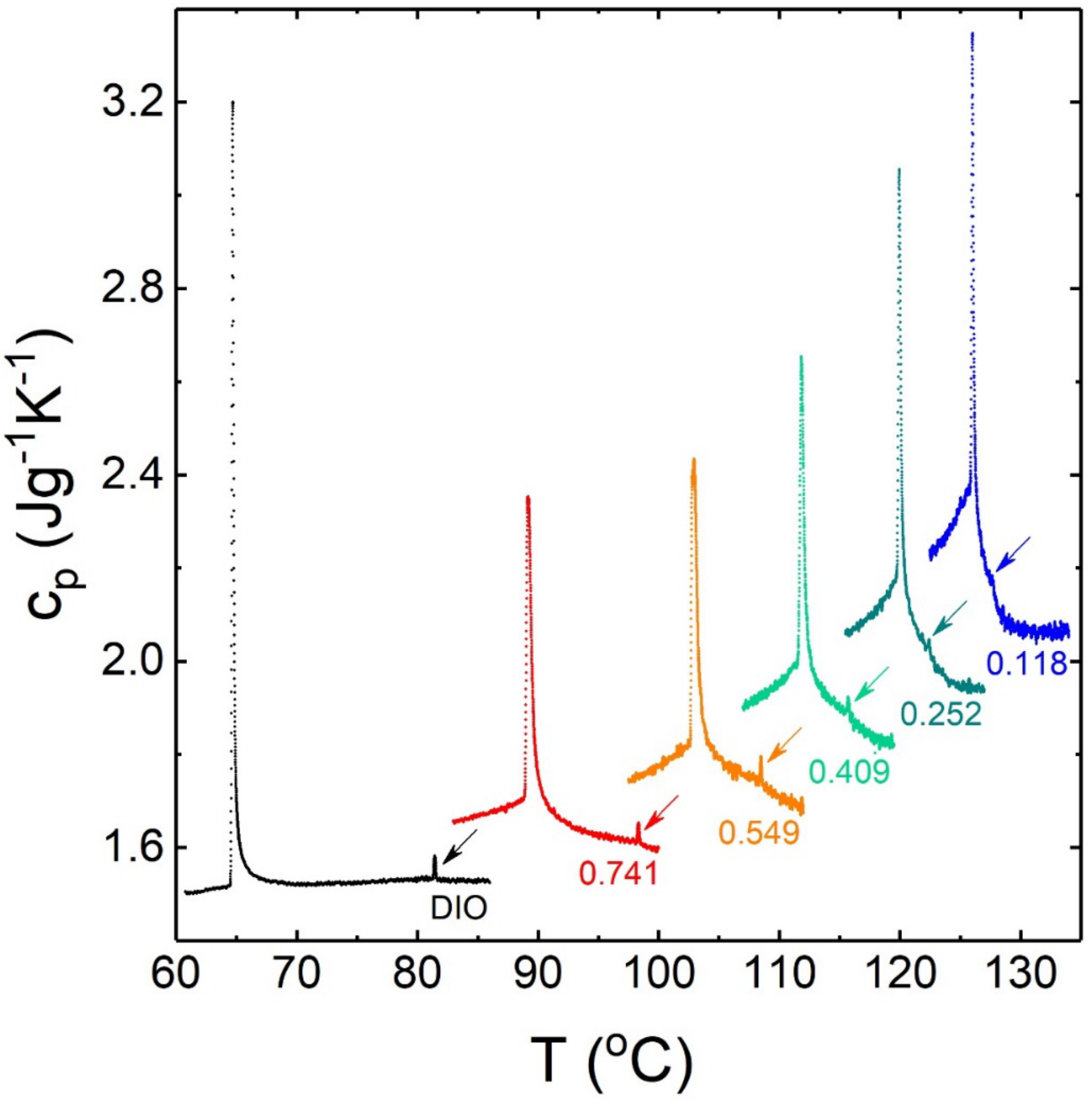

Fig.7 (Color online). Temperature dependence of the specific heat capacity $c_p$ from well into the ferroelectric nematic phase $N_F$ to well into the nematic phase N, crossing the $N_F$-$N_x$ and the $N_x$-N transitions for DIO and five mixtures with weight fractions $w_D$ of 0.741 (red), 0.549 (orange), 0.409 (green), 0.252 (dark green) and 0.118 (blue), respectively. The arrows point to the very weak signals of the $N_x$-N transitions.

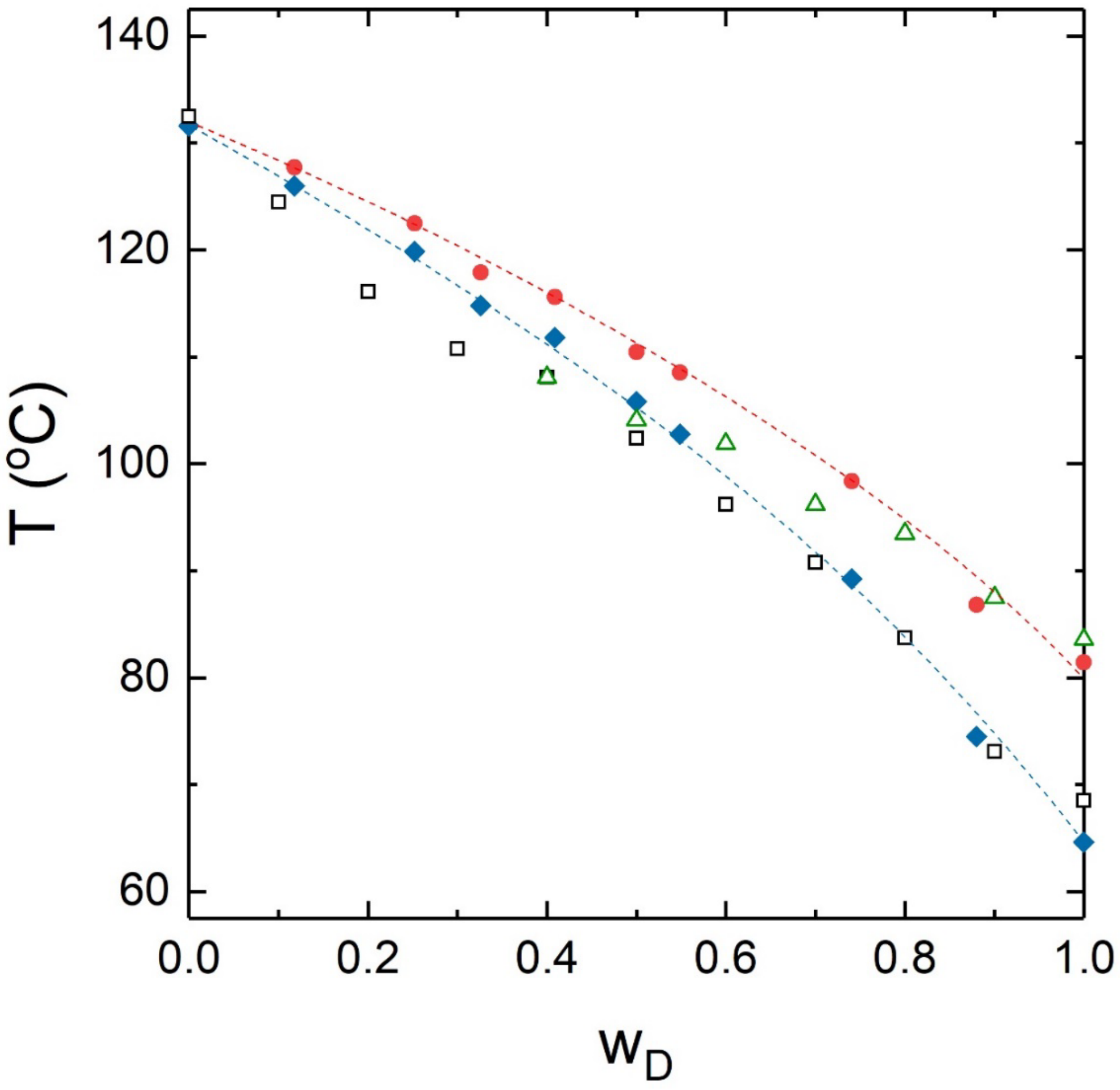

Fig. 8. (Color online) Phase diagram for several mixtures of RM734+DIO. The red solid circles are present ASC values for the $N_x$-N transition temperatures. The solid blue rhombs are the ASC results for the $N_F$-$N_x$ transition temperatures for DIO and the mixtures plus the data point at $w_D = 0$ for RM734 from Ref. [26]. The dashed red and blue curves represent quadratic trendlines through the data. The green open triangle and open black square symbols represent values derived from Figure 2 in reference [25]. The triangles are, according to Ref. [25], for the $N_x$-N transitions and the squares are for the $N_F$-$N_x$ transition for $w_D$ >0.4 and for the $N_F$-N transition for $w_D$ < 0.4.

*3.2.2 True latent heats at the transitions and width of the two-phase regions*

Detailed analysis of $h(T)$ in combination with the corresponding $c_p(T)$ has revealed small two-phase regions for both transitions of all mixtures and discontinuous enthalpy changes, indicating an albeit very weakly first-order character of the transitions. Fig. 9 gives an overview of the widths of the two-phase regions and Fig. 10 of the true latent heat values $L_{N_F\text{-}N_x}$ (J/g) and $L_{N_x\text{-}N}$ (J/g) for the $N_F$- $N_x$ transition (upper blue data points) and for the $N_x$-N transition (lower red data points) .The latent heats for the $N_x$-N transitions are about a factor of 20 smaller than the ones for the $N_F$-$N_x$ transitions. The dotted lines in the figure are quadratic trend lines. The data point for $N_x$-N transition of neat RM734 is from Ref. [26].

As already shown in Fig. 5 for neat DIO, the enthalpy change associated with the transition between the $N_x$ and the N phases is extremely small. The $c_p(T)$ data in Fig. 5 indicate the absence of pretransitional effects, making this transition purely, but very weakly, first order. The same characteristics were observed for the $N_x$-N transitions in the eight investigated binary mixtures of RM734 and DIO. However, with decreasing DIO concentration this transition shifts to a temperature closer to that of the $N_F$-$N_x$ transition and also exhibited smaller peaks, becoming barely visible with decreasing $w_D$ on the pretransitional high-temperature $c_p(T)$ slope of the latter transition. Moreover, the corresponding widths and the latent heats gradually decreased with decreasing $w_D$. For the lowest $w_D = 0.118$, a latent heat value of only 0.002 ± 0. 002 J/g was obtained.

The $N_F$-$N_x$ transition in the mixtures is also first order with decreasing latent heats upon decreasing $w_D$ from the DIO value of 0.23 ± 0.03 J/g to the value of 0.115 ± 0.005 J/g for RM734 [26]. As in the case of DIO and RM734, the mixtures also exhibit substantial, fluctuations induced, pretransitional specific heat capacity changes near the $N_F$-$N_x$ transitions, in particular in the high-temperature phases. In Fig. 7 it can be observed that the $c_p(T)$ variation extends well into the N phase above the $N_x$-N transition. In fact, because of its smallness, the $c_p(T)$ peak associated with the $N_x$-N transition, has a negligible effect on the high-temperature $c_p(T)$ behavior associated with the $N_F$-$N_x$ transition.

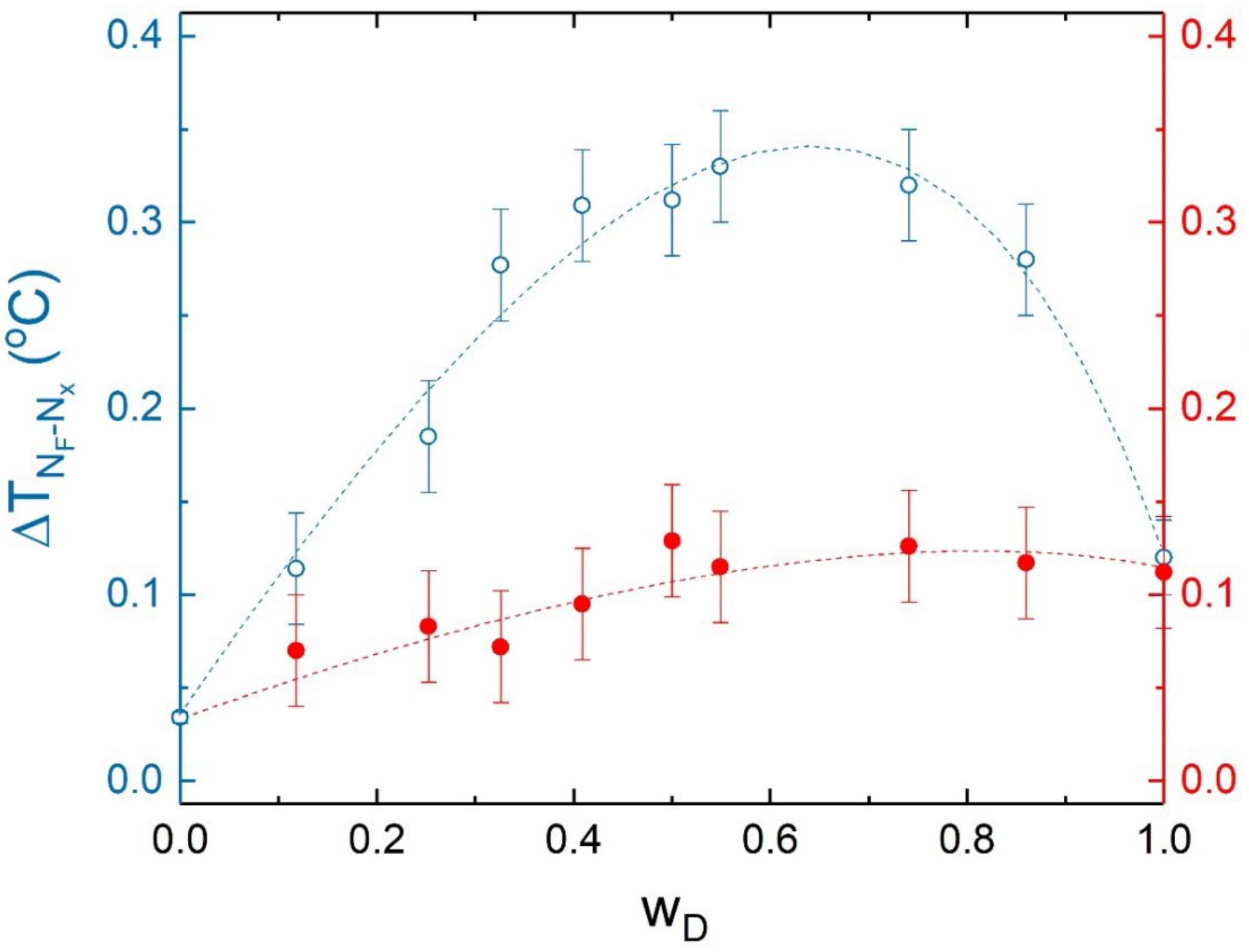

Fig. 9. (Color online) Overview of the observed widths of the two-phase regions for neat DIO and RM734 and for the eight mixtures investigated. The upper blue (open circles) data points denote the $N_F$-$N_x$ transition. The blue data point at $w_D = 0$ is the value reported in Ref. [26] for the $N_F$-N transition for RM734. The blue dashed line corresponds to a fit with a cubic equation. The lower red (solid circles) data points denote the $N_x$-N transition and the red dashed line is from a quadratic fit. At $w_D = 0$ no red data point is plotted because no $N_x$-N transition was reported in Ref. [26].

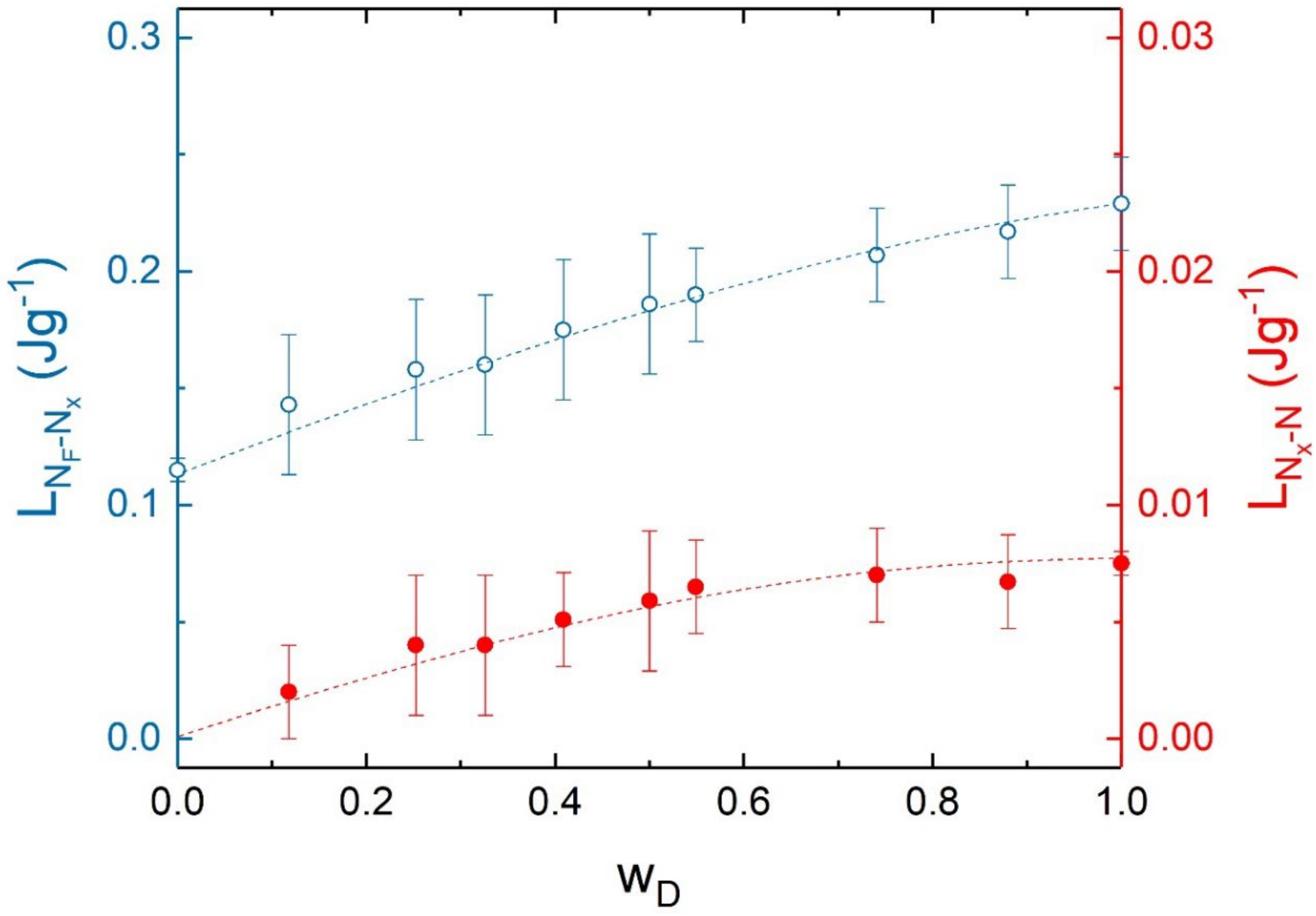

Fig. 10. (Color online) True latent heat values $L_{N_F\text{-}N_x}$ and $L_{N_x\text{-}N}$ for the $N_F$-$N_x$ . transition (upper blue data points) and for the $N_x$-N transition (lower solid red data points). Note that different scales are used among the right and left vertical axis. Thus the latent heats for the $N_x$-N transitions are about a factor of 20 smaller than the ones for the $N_F$-$N_x$ transitions. The dashed lines are quadratic trend lines. The blue data point at $w_D = 0$ is the value reported in Ref. [26] for the $N_F$-N transition of RM734. At $w_D = 0$ no red data point is plotted because no $N_x$-N transition was reported in Ref. [26].

### *3.2.3 Anomalous pretransitional specific heat capacity behavior at the $N_F$-$N_x$* transitions

As already pointed out for neat DIO in section 3.1.4 the fact that both $c_p(T)$ as well as $h(T)$ were obtained by ASC, offers a specific and unique way of investigating pretransitional behavior by combining the specific heat capacity data $c_p(T)$ with the quantity $C$, derived from $h(T)$, and introduced in Eq. (8). As explained in section 3.1.4 the method is in principle intended for second-order phase transitions, but can be extended to very weakly first-order phase transitions. This approach allows direct access to experimental values for the effective critical exponent α, introduced in the power law expression of Eq. (4). This method was successfully used for RM734 [26] and for DIO above in Section 3.1.4 (see also Fig. 6). Since the $N_F$-$N_x$ latent heats for the mixtures are as small as or smaller than for DIO (see Fig. 10), it can be expected that the method to be applicable for the mixtures. The increased two-phase width (see Fig. 10) of the mixtures is a concern, because two-phase region data have to be excluded from the analysis. In Fig. 11, an overview of log-log plots based on Eq. (11) and similar to Fig. 6 for DIO, is presented. As already pointed out in section 3.1.4 the slope of these curves correspond with the effective critical exponent $\alpha$ values. It can be observed that with decreasing $w_D$ the slope decreases from the DIO value (around 0.9) to the RM734 value (around 0.5 [26]). The curves on the low temperature side of the transition exhibit (much) larger uncertainty because the critical amplitude (see Eq. (4)) for that phase is about 2.5 times smaller than those that were observed on the high temperature side. Fig. 12 gives an overview of the resulting values for the effective critical exponent $\alpha$ as a function of weight fraction $w_D$ of DIO. Within the estimated uncertainties, an almost linear $w_D$ dependence between the DIO and RM734 values can be observed.

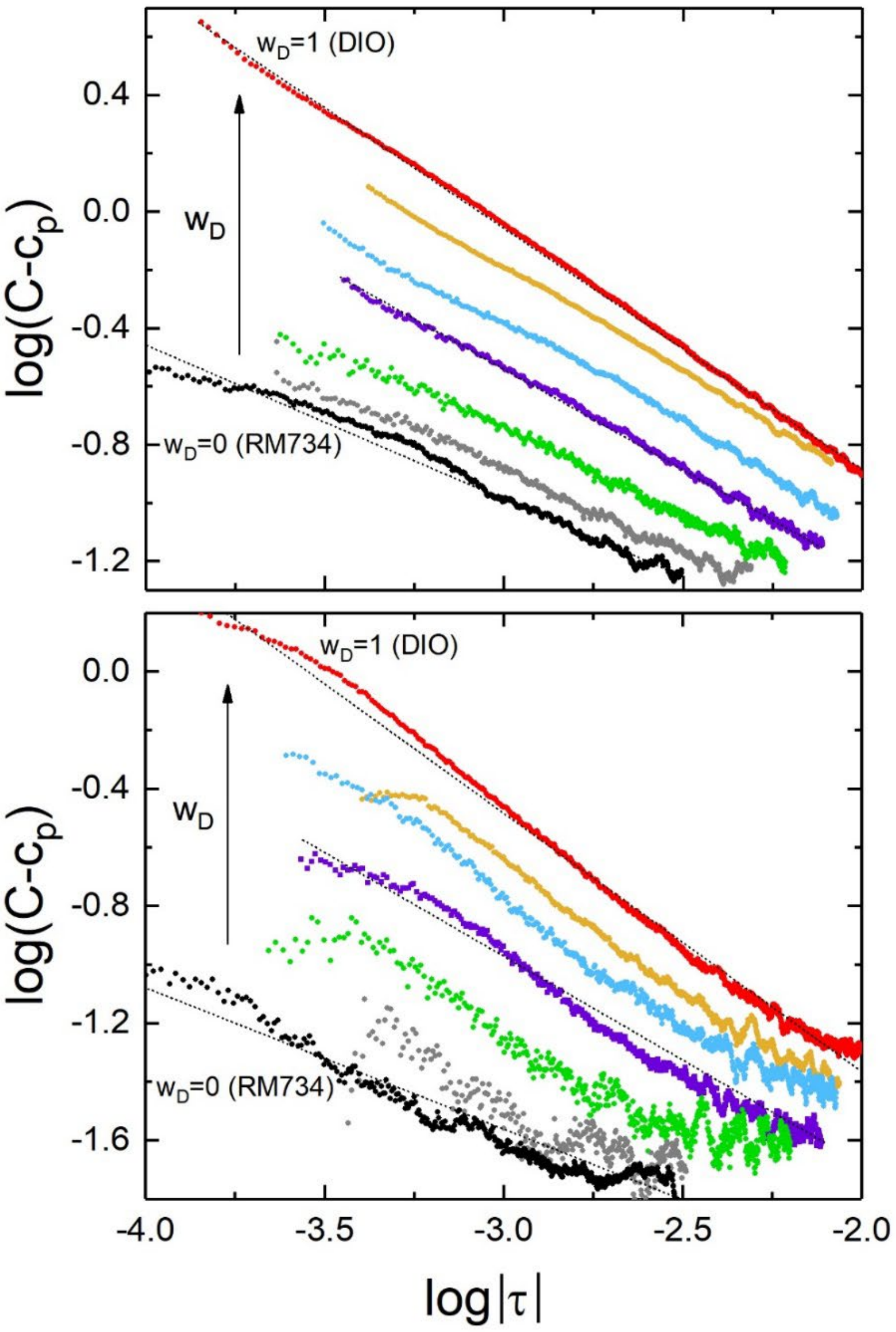

Fig. 11. (Color online). Double logarithmic plot for the difference (C-cp), expressed in J/(gK) as a function of the reduced temperature difference $\tau$ (see Eq. 8). The upper panel shows data above the transition temperature $T_{tr}$. The lower panel shows data below $T_{tr}$. Red data points are for neat DIO and the black ones for neat RM734. The intermediate curves are for mixtures of DIO and RM734 with weight fraction $w_D = 0.741$ (oranje), $w_D = 0.549$ (blue), $w_D = 0.409$ (violet), $w_D = 0.252$ (green), $w_D = 0.118$ (gray). Some of the data sets have been shifted upward for display reasons. The slopes of the curves correspond to the effective critical exponent $\alpha$ of Eq. (8).

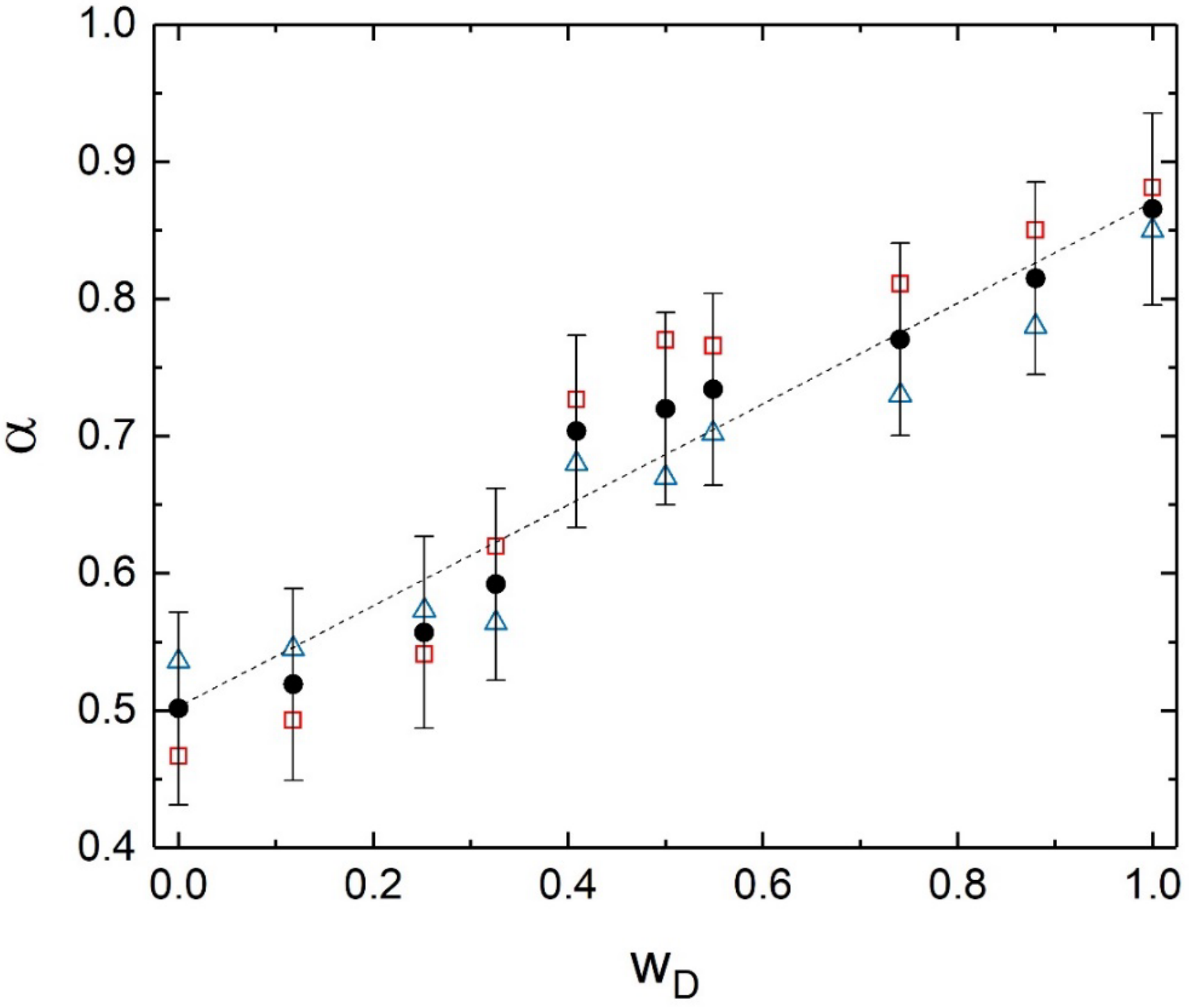

Fig. 12. (Color online). Comparisons, as a function of the weight fraction of DIO, between the effective critical exponent $\alpha$ values (see Eq. (4)) at the $N_F$-$N_x$ transition for neat RM734 (to the left) and neat DIO (to the right) and for eight different mixtures of these compounds. The open (red) squares are from $c_p(T)$ data above the transitions and the open (blue) triangles below the transition. The solid (black) circles are average values and the dashed line shows a linear fit through these data.

## 4. DISCUSSION

As already mentioned in the introduction, more than a century ago Debye [5] and Born[6] discussed the possibility of a ferronematic phase by extending the ferromagnetic Weiss model to molecular electric dipoles, but only recently such a phase behavior was observed experimentally. In 2017 two different molecules were reported to exhibit ferroelectric like features independently by two different groups. One molecule was RM734 [7-12] and the other one DIO [8]. In the meantime, many more compounds from both the RM734 family [15] and the DIO family [20] have been reported to exhibit nematic ferroelectricity. The fact that RM734 and DIO are members of separate molecular families with quite different molecular structures, makes ferronematic behavior a surprise at first sight. This has motivated investigations of differences and similarities in phase behavior of these molecules and of their mixtures [23,25].

In the past high-resolution calorimetric studies have played a significant role in providing information on energy effects near many liquid crystal phase transitions [27-33,42,43]. In fact, the many different phases and phase transitions make liquid crystals excellent model systems for testing general phase transition and critical phenomena concepts. The first-order or the second-order (or continuous) nature of transitions and their critical exponents and universality class have been investigated extensively. Adiabatic scanning calorimetry has proven to be an important tool to discriminate between first-order and second-order transitions and render high-resolution information on pretransitional heat capacity behavior.

A previous paper [26] presented the results of a detailed ASC investigation of the ferroelectric nematic to nematic transition ($N_F$-N) transition of RM734. As already pointed out in the introduction the transition was proven to be very weakly first order with a true latent heat $L = 0.115 \pm 0.005$ J/g. The pretransional specific heat capacity behavior in the high-temperature nematic phase was found to be substantially larger than in the low temperature ferronematic phase. For most phase transitions the opposite is common. Analysis of this pretransional behavior resulted in a critical exponent (see Eq.7) $\alpha = 0.50 \pm 0.05$. This value of the critical exponent suggested, although, first-order, the transition to be close to a tricritical point.

From the new results on DIO, presented in Section 3.1, some similarities and differences with RM734 can be observed. As already established, an important difference with RM734 is the presence of the $N_x$ phase between the $N_F$ and the N phases and the presence of an additional $N_x$-N phase transition. This transition is weakly first order, with a latent heat $L = 0.0075 \pm 0.0005$ J/g and exhibits no pretransitional effects (see Fig. 5). The temperature dependence of $c_p(T)$ and $h(T)$ around the $N_F$-$N_x$ transition (see Fig. 3 and 4) is similar to that of the $N_F$-N transition of RM734. The $N_F$-$N_x$ transition of DIO is also weakly first order, albeit with a larger latent heat $L = 0.23 \pm 0.03$ J/g which is about twice the one of RM734. Power law analysis (see section 3.1.4 and Fig. 6) of the pretransional specific heat capacity resulted in an effective critical exponent $\alpha = 0.88 \pm 0.10$. This is an unusual value which is quite different from the $\alpha = 0.50 \pm 0.05$ observed for RM734. This, likely being a crossover value, cannot obviously be linked to a specific type of universality class.

In a further effort to find insight in the similarities and differences in the phase transition behavior of DIO and RM734, we carried out ASC runs for eight mixtures of these compounds. From these measurements and in the phase diagram in Fig. 8 we did not find any evidence of phase separation, thus confirming the complete miscibility reported by X. Chen *et al.* [25]. However, from the comparison in Fig. 8 between our phase transitions temperatures and the ones from Ref. [25], two significant differences are observed. First, in our case the $N_x$ phase persists over the entire concentration range (at least from $w_D = 0.12$ to 1.0) and does not disappear for weight fractions below $w_D \approx 0.40$ as was reported in [25]. Second, our data show curvature (see the quadratic fits in Fig. 8), deviating from the linearity that was presumed in reference [25]. This discrepancy, and the quoted linearity, is most likely the result of the lower resolution of the transition temperatures determination in Ref. [25]. Chen *et al.* [25] tested ideal mixing of the paraelectric and ferroelectric phases on the basis of an expression derived by Van Hecke [44] on the basis of the so-called equal G analysis. In this approach the total Gibbs energies between two phases in equilibrium are set equal to describe phase coexistence. Ideal as well as nonideal systems under constant pressure have been considered. For a two-component ideal mixture (absence of excess Gibbs energies) and provided that the heat capacity effects are small compared to transition entropies, the following expression was obtained [44] for the transition temperatures as a function of the mole fraction x of component 2.

$$T(x) = \frac{(1-x)\Delta S_1 T_1 + x\Delta S_2 T_2}{(1-x)\Delta S_1 + x\Delta S_2}. \quad (12)$$

$T_1$ and $T_2$ are the transition temperatures of the pure components and $\Delta S_1$ and $\Delta S_2$ are the corresponding (constant) transition entropies. Provided the values of $T_1$ and $T_2$ of the pure compounds are known, the number of unknown parameters on the right hand side of this equation can be reduced to one, allowing for a single parameter fit to experimental data of the transition temperatures of the mixtures:

$$T(x) = \frac{(1-x)T_1 + xsT_2}{(1-x)+xs}, \quad (13)$$

with $s = \Delta S_2/\Delta S_1$ the fitting parameter. From Eq. (13) it can immediatly be observed that $T(x)$ (with x the mole fraction of component 2) is only linear for s=1. For values of s > 1 the curve $T(x)$ is convex-shaped and for s < 1 it is concave-shaped. However, this does not indicate for the system to be non-ideal [44].

In order to test whether Eq. (13) is applicable to our $N_F$-$N_x$ phase transition data, we have converted the data from a $T(w_D)$ representation, with $w_D$ the weight fraction of DIO, as in Fig. 8, to $T(x)$, with $x$ the mole fraction of DIO, i.e., compound number 2, in Eq. (13) using the molecular weights $MW_{RM734}$ = 423 g/mol and $MW_{DIO}$ = 510 g/mol. Fig. 13 gives the $N_F$-$N_x$ and the $N_x$-N transition temperature data versus the mole fraction of DIO. Before using Eq. (13) in the data analysis, we can consider some boundary conditions for ideal systems on the slopes of the coexistence curve at $x = 0$ and $x = 1$. The following two slope equations apply [44]:

$$D_0 \equiv \left.\frac{\partial T}{\partial x}\right|_{x\to 0} = -s(T_1 - T_2), \quad (14)$$

$$D_1 \equiv \left.\frac{\partial T}{\partial x}\right|_{x\to 1} = (T_2 - T_1)/s. \quad (15)$$

Making the product of the Eqs. (14) and (15) results in:

$$D_{0,1} \equiv \left.\frac{\partial T}{\partial x}\right|_0 \left.\frac{\partial T}{\partial x}\right|_1 = (T_2 - T_1)^2 \quad (16)$$

In order to arrive at initial slope values at $x = 0$ and $x = 1$ for the $N_F$-$N_x$ transitions we have used the following very good fitting result $T(x) = -19.83x^2 - 47.2x + 404$ (see Fig. 13). Making the derivative resulted in $D_0 = -47.2 \pm 0.2$ K and $D_1 = -87 \pm 1$ K. The product $D_{0,1} = D_0D_1 = 4098 \pm 65$ $K^2$ compares reasonably well with $(T_2 - T_1)^2 = (67 \pm 1)^2 = 4489 \pm 135$ $K^2$.

Eqs. (14) and (15) allowed also to arrive at values of $s = \Delta S_2/\Delta S_1$. From $D_0$ we obtain $s = 0.71 \pm 0.02$ and from $D_1$ we get $s = 0.77 \pm 0.02$ . Fitting the transition temperatures with $s$ as adjustable parameter in Eq. (13) results in $s = 0.78 \pm 0.03$.

The above values can be compared with the experimental ratio $s = \Delta S_{DIO}/\Delta S_{RM734}$ which can either be calculated from the true latent heat values $L$ or from $\Delta h_{tr}$ which have been reported in Section 3.1.2. Using only the true latent heat values resulted in $s = 2.00$, which differs by a large amount from the above values. However, using the (total) $\Delta h_{tr}$ values of section 3.1.2 yields $s = \Delta S_{DIO}/\Delta S_{RM734} = 0.73/1.07 = 0.68 \pm 0.12$, which is close to the slope results discussed above. This indicates that it is necessary to include in the ΔS the pretransition fluctuation contribution to be consistent with the results obtained by the analysis using Eqs. (12-16). The comparability of the $s$ value seems to confirm the proposed ideality [25] of the binary DIO+RM734 system, albeit with a nonlinear $T(x)$ curve.

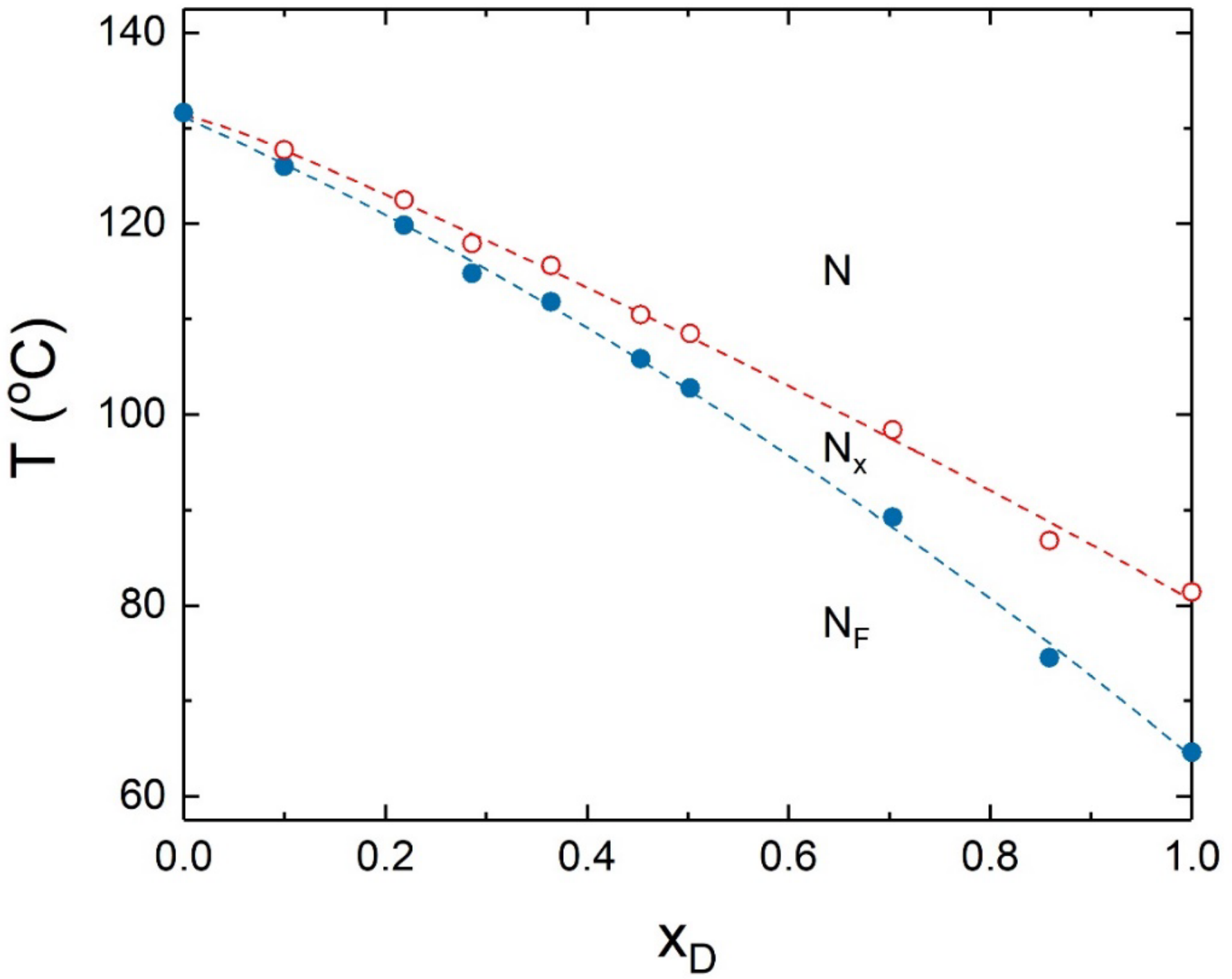

Fig. 13. (Color online) Transition temperatures as a function of the mole fraction of DIO . The red open circles are data of the $N_x$-N transition temperatures, and the filled blue circles are the results for the $N_F$-$N_x$ transition temperatures of DIO and of the mixtures and for the $N_F$-N transition

temperature of RM734 (at x = 0) ,reported in Ref. [26]. The dashed red and blue curves represent quadratic trendlines through the corresponding data points.

The procedure established in section 3.1.2 to arrive at $\Delta h_{tr}$ for the neat DIO and RM734 can be applied to the mixtures as well. The results of this analysis are summarized in Fig. 14. An almost linear weight fraction $w_D$ dependence can be observed, which is decreasing with $w_D$, which is opposite to what can be seen in Fig. 9 for the latent heat.

Eqs. (13) to (16) have also been applied to the transition temperature data for the $N_x$-N transition. For the slope analysis at x = 0 and x = 1 we used the fit equation $T(x) = -8.46x^2 - 42.9x - 405$ (see Fig. 13). The derivatives resulted in $D_0 = -42.9 \pm 0.3$ and $D_1 = -60 \pm 1$. From $T_2 - T_1 = 50 \pm 1$ K and $D_0$ we obtain $s = 0.86 \pm 0.03$ and from $D_1$ we get $s = 0.83$. The product $D_{0,1} = D_0D_1 = 2567 \pm 70$ $K^2$ compares well with $(T_2 - T_1)^2 = (50.0)^2 = 2500 \pm 100$ $K^2$. Fitting the transition temperatures with $s$ as an adjustable parameter in Eq. (13) results in $s = 0.88$. In this case comparison with the experimental ratio $s = \Delta S_{DIO}/\Delta S_{RM734}$ was not feasible, because no $N_x$-N transition has been detected [26] in compound 1 (RM734).

Detailed analysis (see section 3.2.2) of the specific enthalpy $h(T)$ data in combination with the specific heat capacity $c_p(T)$ data confirmed the first-order character of both the $N_F$-$N_x$ and the $N_x$ -N transitions in all the investigated mixtures. While the two-phase region for $N_x$ -N gradually decreased with decreasing weight fraction of DIO, substantially wider two-phase regions were observed in the range of $w_D$ around 0.5 for the $N_F$-$N_x$ transitions (see Fig. 9). However, for both type of transitions the true latent heats $L$ gradually decreased with decreasing $w_D$ (see Fig.10). Power law analysis for the mixtures, similar to the ones performed for neat RM734 and DIO, resulted in a set effective critical exponent $\alpha$ values intermediate between those of RM734 ($\alpha$ =0.50 ± 0.05) and of DIO ($\alpha$ =0.88 ± 0.10), with an almost linear dependence on $w_D$ (see Fig. 12). As the RM734 value could be associated with the possibility of a nearby tricritical point [26], the DIO value and the evolution with $w_D$ cannot obviously be linked (to cross-over) to a specific universality class.

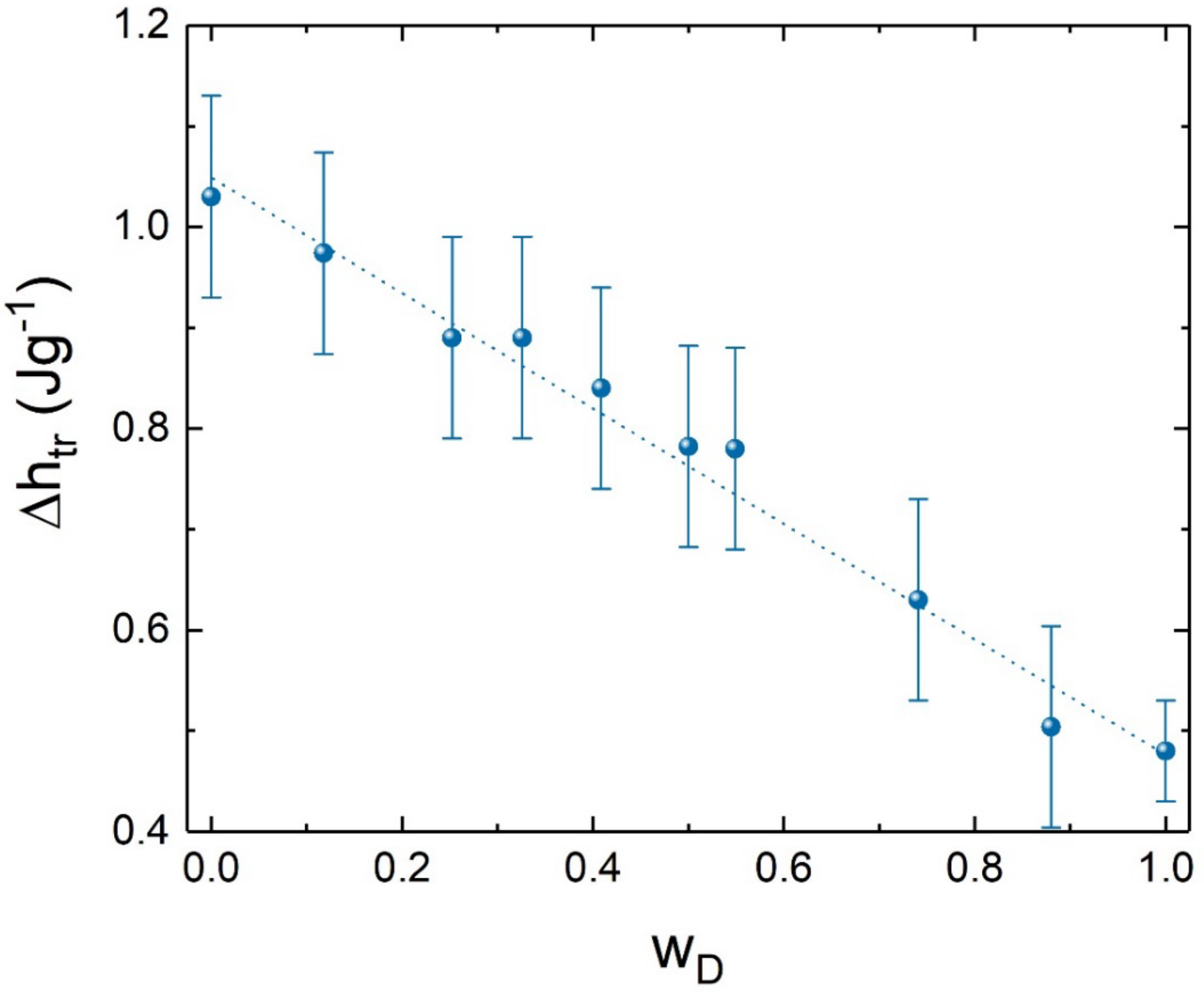

Fig. 14 Total specific enthalpy change $\Delta h_{tr}$ associated with the transitions from the $N_F$-$N_x$ transitions as a function of the weight fraction $w_D$ of DIO (see Eq. (3)). The point at $w_D = 0$ is for the $N_F$-N transition of RM734 [26].

## 5. SUMMARY AND CONCLUSIONS

High-resolution adiabatic scanning calorimetry (ASC) has been used to obtain simultaneously the temperature dependence of the specific enthalpy $h(T)$ on the specific heat capacity $c_p(T)$ of the compound 2,3',4',5'-tetrafluoro[1,1'-biphenyl]-4-yl-2,6-difluoro-4-(5-propyl-1,3-dioxan-2-yl) benzoate (DIO) and on several of its mixtures with 4-[(4-nitrophenoxy)carbonyl]phenyl 2,4-dimethoxybenzoate (RM734). Both compounds exhibit a low-temperature ferroelectric nematic phase ($N_F$) and a high-temperature nematic phase (N). However, in DIO these two phases are separated by an intermediate phase ($N_x$). From the detailed data of $h(T)$ and $c_p(T)$, we found that the intermediate $N_x$ phase was present in all the mixtures over the complete composition range, albeit with strongly decreasing temperature width for that phase with decreasing mole fraction of DIO ($x_{DIO}$). The $x_{DIO}$ dependence of the transition temperatures for both the transitions could be well described by a quadratic function. Both these transitions were weakly first-order. The true latent heat of the $N_x$-N transition of DIO was as low as $L = 0.0075 \pm 0.0005$ J/g and was $L = 0.23 \pm 0.03$ J/g for the $N_F$-$N_x$ transition, which is about twice the previously reported value of 0.115 J/g for the $N_F$-N transition in RM734. In the mixtures both transition latent heats decrease gradually with decreasing $x_{DIO}$. At all the $N_x$-N transitions pretransition fluctuation effects are absent and these transitions are purely though very weakly first order. As observed in RM734 the transition from the $N_F$ to the higher temperature phase exhibits substantial pretransitional behavior, in particular in the high-temperature phase. Power law analysis of $c_p(T)$ resulted in an effective critical exponent $\alpha = 0.88 \pm 0.1$ for DIO. In the mixtures the effective $\alpha$ value decreased with decreasing $x_{DIO}$ to $\alpha = 0.50 \pm 0.05$ the value reported for RM734 [26]. Ideal mixture analysis [44] of the phase diagram was consistent with ideal mixture behavior provided the total transition enthalpy change was used in the analysis.

## Acknowledgement

G.C. acknowledges financial support from Project P1-0125 of the Slovene Research Agency. W.J. acknowledges the China Scholarship Council for a PhD scholarship and GHM funding by Diamond Light Source (UK) through project SM30755.